\documentclass[%
 reprint,
superscriptaddress,
groupedaddress,
preprintnumbers,
nofootinbib,
 amsmath,amssymb,
 aps,
dvipsnames
]{revtex4-2}
\pdfoutput=1

\usepackage{graphicx}
\usepackage{dcolumn}
\usepackage{bm}
\usepackage[normalem]{ulem} 
\usepackage{tikz}
\usetikzlibrary{decorations.pathmorphing}
\usetikzlibrary{decorations.markings}
\usetikzlibrary{calc}
\usepackage{verbatim}
\usepackage{caption}
\usepackage{subcaption}
\usepackage{mathrsfs}

\newcommand{\de}{\delta}
\newcommand{\mP}{\mathscr{P}}
\newcommand{\fD}{\mathscr{D}}
\newcommand{\Gphi}{\boldsymbol{\Gamma}}
\newcommand{\CdT}[1]{\mathcal{T}_{ #1}}

\begin{document}
\preprint{IPPP/22/44}
\title{On the effective action for scalars in a general manifold to any loop order}
\author{Rodrigo Alonso}
\author{Mia West} 
\affiliation{Institute for Particle Physics Phenomenology, Durham University, South Road, Durham, DH1 3LE}

\begin{abstract}
	Functional methods and a derivative expansion are employed for laying out a procedure to compute the effective action  to any loop order, for scalar fields parametrising an arbitrary Riemannian manifold, while maintaining explicit field-space covariance.  These results are of use in the characterization of effective field theories for electroweak symmetry breaking and extend a geometric perspective in field space beyond one loop.
\end{abstract}

\maketitle


\section{Introduction}

Effective field theories (EFT) provide a model-independent theory framework to explore energy frontiers. This exploration has, in the past, yielded predictive models which improve our understanding of Nature. Inspired by the success of models one can try and use them as exploration tools yet it is useful to keep in mind that, compared with EFT, what they offer in predictivity they lack in generality. This discussion finds meaning in the most prominent energy frontier of our era, the electro-weak scale, presently being explored at the LHC.  The importance of this endeavour can hardly be overstated; it holds the answer to the mechanism behind gauge symmetry breaking, the generation of masses and the hierarchy problem. In our search for new physics at LHC, EFT has been gaining prominence given the experimental absence of predictions from postulated models thus far. The honing of EFT as a tool to describe electro-weak physics has produced Higgs Effective Field Theory (HEFT) as the most general EFT to describe Lorentz and gauge invariant model independent processes at LHC. A subset of this theory space is known as the Standard Model EFT (SMEFT) and presents a linear realization of EWSB. A dichotomy has hence formed; SMEFT or HEFT/SMEFT, this latter referred here as non-linear theory space or non-linear theories. The exploration of this non-linear theory space has taken theory into new territory in subjects such as geometry for field space \cite{Alonso:2016oah,Alonso:2015fsp}, the UV completion of non-linear theories \cite{Falkowski:2019tft,Cohen:2021ucp,Alonso:2021rac}, and amplitude methods  \cite{Cohen:2021ucp,Cheung:2021yog,Cheung:2022vnd,Cohen:2022uuw}; the phenomenology in turn has been studied in e.g.~\cite{Brivio:2013pma,deBlas:2018tjm,Gomez-Ambrosio:2022qsi}. This letter is concerned with the geometric description and quantum corrections. 

Differential geometry proves useful when describing scalars parametrising a general manifold, the scalar sector of the electro-weak theory being a current case study just as pions were in the past; it connects tensors in field space with physical quantities and preserves field reparametrization invariance along with symmetry. An appropriate tool to preserve these qualities all throughout derivation of results are functional methods. Functional techniques, referred to as the background field method during their inception in particle physics \cite{DeWitt:1967ub,Honerkamp:1971sh,tHooft:1973bhk,Kallosh:1974yh}, have been in use for half a century in effective potential computations \cite{Coleman:1973jx,Ford:1992pn,Ghilencea:2016ckm}, gravity \cite{DeWitt:1967ub,Vilkovisky:1984st}, string theory \cite{Friedan:1980jf,Alvarez-Gaume:1981exa} and gauge invariant computations \cite{Abbott:1980hw,Jack:1982hf,Bornsen:2002hh,Ivanov:2022aco} to name a few. Often these methods implement computations with the heat-kernel technique; an alternative yet less developed option being known as covariant derivative expansion (CDE) \cite{Chan:1985ny,Gaillard:1985uh,Henning:2014wua,Alonso:2019mok}. 
This work will employ functional methods and a CDE to lay out the extension of quantum loop computations beyond one loop in the geometric description, including gauge and gravitational interactions, together with a sample two loop computation.

As means of introduction and for later reference, let us sketch here the functional method formulation to the extent in which is most commonly encountered, i.e. with just sufficient depth to derive one loop results for scalar fields in a flat manifold.
The partition function and generating functional read, in terms of the tree level action,
\begin{align}\label{Bgng}
Z[J]=\int [d\phi] \,e^{i(S[\phi]+J\cdot\phi)/\hbar}\equiv e^{iW[J]/\hbar}\,,
\end{align}
where $\hbar$ is Planck's constant that we set to $1$ until further notice, $[d\phi]$ is the functional measure and we use $\cdot$ for the scalar functional product, also denoted with DeWitt notation (e.g. $\phi^x$), as 
\begin{align}\label{defindx}
J\cdot \phi\equiv\int d^dx\, \phi(x) \,J(x)\,\equiv J_x\,\phi^x\,
\end{align}
where $d$ is the space-time dimension.
Next the Legendre-transform gives the effective action 
\begin{align}
&\Gamma[\varphi]\equiv W[J(\varphi)]-\varphi\cdot J(\varphi)\,,&
&\varphi\equiv\frac{\de W}{\de J}\,,
\end{align}
where one would use the definition of $\varphi$ to invert and find $J(\varphi)$, to be substituted in the definition of $\Gamma$ above. 

The n-point scattering matrix $\mathcal{S}$ follows from the poles of correlation functions as
\begin{align}
   \mathcal S_{(n)}=&\frac{1}{Z[0]}\left(\prod_i^n\frac{p_i^2-m_i^2}{\sqrt{\mathcal Z}}\frac{\de}{i\de J(p_i)}\right)Z[J]\\
   =&\left(\prod_i\frac{p_i^2-m_i^2}{\sqrt{\mathcal Z}}\right) \left(\varphi+i\frac{\de^2\Gamma}{\de\varphi^2}\frac{\de}{\de\varphi}\right)^n
\end{align}
with $\mathcal Z$ the field normalization (i.e. the residue of the two point function $\langle \phi\phi\rangle =i\mathcal Z/(p^2-m^2)+\dots$) and $\delta/\delta\varphi$ is the functional deriviative, e.g.
\begin{align}
\frac{\de\phi^x}{\de\phi^y}&=\de^d(x-y)\,,\\ \frac{\de S}{\de \phi^y}&=\int d^4x \frac{\de\phi^x}{\de\phi^y}\left(\frac{\partial \mathcal L}{\partial\phi}-\partial_\mu\frac{\partial \mathcal L}{\partial\partial_\mu\phi}+\dots\right)\,,
\end{align}
Given the inability to carry out the path integration in general, one resorts to an expansion on radiative corrections with $\phi=\phi_0+\phi_{\rm q}$, $\phi_0$ the background field, and $\phi_{\rm q}$ parametrises virtual fluctuations, i.e. it is the dummy variable for the path integration and as such not featuring in the final results. The expansion reads,
\begin{align}\label{ExPI}
	S[\phi]+J\cdot\phi=S[\phi_0]+J\cdot \phi_0 
	+\sum_{n=2}\frac{\phi_{\rm q}^n}{n!}\frac{\de^nS}{\de\phi^n}[\phi_0]\,,
\end{align}
 and the field $\phi_0$ is defined to cancel the linear term in $\phi_{\rm q}$
\begin{align}
\frac{\de S}{\de \phi}[\phi_0]+J\equiv0\,,
\end{align}
and is hence a function of the source $J$.
The first term in this expansion, $S[\phi_0]+J\cdot\phi_0$, is a constant factor which can be pulled out of the path integral, giving the classical generating functional and effective action: $\Gamma^{(0)}[\varphi]=S[\varphi]$ and $\Gamma^{(n)}$ is the n'th order quantum correction, $\Gamma=\Sigma_n \Gamma^{(n)}$. Further, the first quantum corrections can be obtained from the next term in the expansion in eq.~(\ref{ExPI}) by performing a Gaussian integral:
\begin{align}
\int [d\phi_q] e^{i\phi_q ^2\de^2\!S[\phi_0]/2}= \frac{
\mathcal N}{\sqrt{\det(-\de^2S)}}\,,
\end{align}
where we have wick rotated to Euclidean space and come back and ${\mathcal N}$ is a constant factor. The effective action at the one loop level reads, after using $\det(A)=e^{{\rm tr}[\log (A)]}$,
\begin{align}\label{oneloop}
\Gamma^{(0)}[\varphi]+\Gamma^{(1)}[\varphi]=S[\varphi]+\frac i2{\rm Tr}\,[\log(-\de^2S[\varphi])]\,,
\end{align}
where the trace is to be taken over space-time and internal indexes in our fields.
 
Sec.~\ref{secTH} presents the extension of this procedure and generalisation of the effective action above to the invariant effective action to arbitrary loop order and scalar manifold, while sec.~\ref{ThreeL} reports the three loop formula and sec.~\ref{TwoLVeff} uses this result to compute the two loop corrections to the effective potential for a scalar manifold of $N$ dimensions and $O(N)$ symmetry. The summary can be found in sec.~\ref{Summ}.


\section{The effective action for any scalar manifold and loop order\label{secTH}}

The abridged version of the functional method derivation of  one loop corrections is the addition of $i$Tr(log($-\de^2S$))$/2$ to the tree-level action. This is a remarkably simple and portable result, yet this simplicity also obscures the generalization to both higher loops and scalar fields spanning a non-trivial manifold $\mathcal M_\phi$, such as pions or other Goldstones where the manifold is a coset,
 $\mathcal M_\phi=\mathcal G/\mathcal H$. This section is concerned with the simultaneous extension in both these directions. One can find in \cite{Jack:1982hf} an accessible account of the extension to higher loops in flat scalar space and a discussion of the formulation for curved field space on \cite{Honerkamp:1971sh}, yet the discussion here presented differs in novel ways: the introduction of covariant correlation functions, the covariant and parallel transport treatment of the source $J$ and a CDE as a technique to evaluate loop corrections including gravitational interactions.   
 
 The method for the generalization is differential geometry and the upshot: the covariantization of the partition function and correlation functions via the introduction of tensors in field space. The derivation of results, presented next, does introduce some notation and geometric concepts which might at times divert from the goal: an invariant effective action as summarized in eq.~(\ref{CR}) and the itemized list just above it, to which the time-pressed reader might skip given it is self contained. 
 
 
\subsection{Covariant correlation functions and invariant partition function}
Consider $n$ scalar fields $\phi^a$, $a=1,...,n$, as a set of `coordinates' parametrizing a Riemannian\footnote{This manifold has a strictly positive definite metric; treating e.g. gauge fields this assumption should be revisited.} manifold $\mathcal M_\phi$. 
 The manifold can be characterized locally by the field-metric $G$ which appears in the kinetic term in the action, $S_{\rm KE}$, and transforms covariantly as made explicit from taking a field transformation $\phi=\phi(\tilde\phi)$, and using the chain rule,
\begin{align}\label{KE}
S_{\rm KE}=&\int d^d x \sqrt{|g|}\frac{1}{2}\partial_\mu \phi^a G_{ab}(\phi)\partial^\mu\phi^b\\
=&\int d^d x\sqrt{|g|} \frac{1}{2}\partial_\mu \tilde\phi^c \frac{\partial \phi^a}{\partial\tilde\phi^c}G_{ab}(\phi(\tilde \phi))\frac{\partial \phi^b}{\partial\tilde\phi^d}\partial^\mu\tilde\phi^d\\
\equiv&\int d^dx \sqrt{|g|}\frac12 \partial_\mu\tilde\phi^c \tilde G_{cd} \partial^\mu\tilde\phi^d\,,
\end{align} 
where $g$ is the space-time metric and $|g|=|\det(g_{\mu\nu})|$.
For consistency and to preserve any symmetries of the system which might be embedded in these field transformations, the expansion should be made covariant and integration over our manifold defined in terms of the (field transformation) invariant measure
\begin{align}
	Z[0]
	{=}\int [d\phi]\sqrt{\det{G(\phi)}}\,e^{iS[\phi]}\,.
\end{align} 

Adding a conventional source term $J\cdot\phi$ does not respect invariance since $\phi$ itself does not transform covariantly. The introduction of the source $J$ is a device to compile correlation functions and these would not be covariant in turn. Let us then explore the generalization to a fully invariant partition function which requires modification of the correlation functions.

For convenience, take the fields to have their origin $\phi^a=0$ at the (tree level) vacuum, performing a shift if necessary, so they describe excitations over the vacuum and $\langle\phi^a\rangle=0$.
In place of $\phi$ which does not transform covariantly, consider Riemann normal coordinates (RNC), $\eta^a$, that follow geodesics. These are specified by the direction in tangent $\phi$-space at which the geodesic curve is departing from zero, and the length along this geodesic, $\sigma$;  the definition is
\begin{align}
\eta^a= \frac{d\phi^a(0)}{d\sigma}\sigma\,,\label{RNCdef}
\end{align}
the point $\phi^a=0$ (i.e. the vacuum) is then $\eta^a=0$ in these coordinates. The connection, or mapping, between the two coordinate systems is given by the solution to the Geodesic equation
\begin{align}
\frac{d\phi^b(\sigma)}{d\sigma} \mathcal D_b\frac{d\phi^a(\sigma)}{d\sigma}=\frac{d^2\phi^a}{d\sigma^2}+\Gphi^a_{bc}\frac{d\phi^b}{d\sigma}\frac{d\phi^c}{d\sigma}=0\,,
\end{align}
with $\Gphi$ the field-connection (or Christoffel symbols) that follows from $G$, and $\mathcal D$ the covariant derivative wrt $\phi$. An expansion around the origin 
\begin{align}
	\phi^a(\sigma)=\sum_{n}\frac{1}{n!}\frac{d^n\phi^a(0)}{d\sigma^n}\sigma^n\,,
\end{align}
put into the geodesic equation returns the higher derivative terms in terms of the tangent vector
\begin{align}
\phi^a=&\sum_{n}\frac{C^a_{i_1\dots i_n}}{n!}\prod_j^n\left(\frac{d\phi^{i_j}(0)}{d\sigma}\sigma\right)\,,
\end{align}
if we now substitute $\eta$ in we obtain a mapping from $\phi$ to RNC $\eta$ coordinates as
\begin{align}
\phi_{[V]}^a=&\sum_n \frac{C^a_{i_1\dots i_n}}{n!}\prod_j^n \eta^{i_j}\\\nonumber
=&\eta^a-\frac{\Gphi^a_{ij}|_V}{2}\eta^i\eta^j-\frac{(\partial_i\Gphi^a_{kl}-2\Gphi^a_{il}\Gphi^l_{jk}|_V}{6}\eta^i\eta^j\eta^k\\&+\mathcal O(\eta^4)\,,\label{mapping}
\end{align}
where the $C$ coefficients are related to the generalized Christoffel symbols, see \cite{Hatzinikitas:2000xe} for higher orders, and $V$ denotes the point of the manifold around which we set up RNC, in particular $V$ marks the vacuum, $\phi^a=0$. From the above definition it follows that both $\eta$ and $\phi$ will excite the same particle out of the vacuum and hence yield the same $S$-matrix elements,
\begin{align}
\langle 0| \phi^c|c,p_\mu\rangle=\langle 0| \eta^c |c,p_\mu\rangle\,,
\end{align}
 yet it is only $\eta$ that transforms covariantly\footnote{Contravariantly to be precise.}. This is most evident in the definition of eq.~(\ref{RNCdef}) in terms of tangent vectors, so it is in terms of these fields that we {\it define} our partition function, and correlation functions:
\begin{align}\label{ZJ}
	Z[J]&=\int [d\eta]\sqrt{\det{\underline G(\eta)}}\,e^{iS[\phi_{[V]}(\eta)]+iJ\cdot\eta}\,,\\
	\mathcal G_{n}(x_1,...,x_n)&\equiv\int [d\eta]\sqrt{\det{\underline G(\eta)}}e^{iS[\phi_{[V]}(\eta)]}\prod_i\eta(x_i),
\end{align}
with
\begin{align}
	\underline G_{ab}=\frac{\partial \phi^c}{\partial \eta^a}G(\phi_{[V]}(\eta))_{cd}\frac{\partial \phi^d}{\partial \eta^b}\,,
\end{align}
 while $J$ is assigned a covariant index, $J_a$. We note that the present definition of the partition function coincides with \cite{Vilkovisky:1984st} up to non-linear terms in the coupling to the source $\mathcal O(\eta^{n\geq 2} J)$ due to the RNC expansion being in this work around the vacuum rather than the mean field as done in ref~\cite{Vilkovisky:1984st}. Having said that, the practical implication of the formulation is both here and in~\cite{Vilkovisky:1984st}: replacing partial derivatives of the action $S$ for covariant derivatives. Under a field transformation $\phi(\tilde\phi)$ which leaves the vacuum at $\tilde\phi=0$, our partition function is invariant and our correlation functions transform as tensors in tangent space
 \begin{align}
 	\tilde Z&=\int [d\tilde\eta]\sqrt{\det{\underline{\tilde G}(\tilde\eta)}}\,e^{iS[\phi(\tilde\phi_{[V]}(\tilde\eta))]+i\tilde J\cdot\tilde\eta} =Z\,,\\
 	\tilde{\mathcal G}_n&=\left(\prod_i\left.\frac{\partial\tilde\phi}{\partial \phi}\right|_{V}\right) \mathcal G_n\,,\label{GnVar}
 \end{align}
 where $\tilde J=(\partial\phi/\partial\tilde\phi ) J$. To keep the discussion explicit let us show the covariant nature of RNC to order $\phi^3$; one has, for each $\phi$ and $\tilde \phi$ coordinates,
 \begin{align}
 \eta^a&=\phi^a+\frac{1}{2}\Gphi^a_{bc}\phi^b\phi^c+\mathcal O (\phi^3)\,,\\
 \tilde\eta^a&=\tilde\phi^a+\frac{1}{2}\tilde\Gphi^a_{bc}\tilde\phi^b\tilde\phi^c+\mathcal O (\tilde\phi^3)\,.
 \end{align}
 A Taylor expansion of the change of coordinates gives to this order,
 \begin{align}
 \tilde\eta^a=&\frac{\partial\tilde\phi^a}{\partial\phi^b}\phi^b+\frac{\partial^2\tilde\phi^a}{2\partial\phi^b\partial\phi^c}\phi^b\phi^c+\frac{\partial \tilde\phi^a }{\partial\phi^d}\frac{\Gphi^d_{bc}}{2}\phi^b\phi^c\\\nonumber
 &+\frac{\partial\tilde\phi^a}{\partial\phi^d}\frac{\partial^2\phi^d}{\partial\tilde\phi^b\partial\tilde\phi^c}\frac12\left(\frac{\partial\tilde\phi}{\partial\phi}\phi\right)^{b}\left(\frac{\partial\tilde\phi}{\partial\phi}\phi\right)^{c}+\mathcal O(\phi^3)\\\nonumber
 =&\frac{\partial\tilde\phi^a}{\partial\phi^d}\left(\phi^d+\frac12\Gphi^d_{bc}\phi^b\phi^c\right)+\mathcal O(\phi^3)=\left.\frac{\partial\tilde\phi^a}{\partial\phi^d}\right|_V\eta^d+ \mathcal O(\phi^3)\,,
 \end{align}
i.e. a covariant transformation around the vacuum.

 To close this section we present a generalisation of the LSZ formula for non-trivial geometry. The matrix $\mathcal S$ is given in terms of correlation functions, which as just described are tensors in field space, so it just field normalization $\mathcal Z$ that remains to be addressed. This relevant step can be made clearer by equating our curved-space kinetic term $G(\partial\phi)^2$ to the action that produces $\langle \phi\phi\rangle =i\mathcal Z/p^2+\dots$, that is $\mathcal Z^{-1} (\partial \phi)^2$, which gives $G|_V=\mathcal Z^{-1}$. The LSZ contains the square root of inverse normalization which corresponds then to the vierbein $\mathbf{e}$
 \begin{align}G_{ab}=\sum_{IJ}\mathbf{e}_a^I\mathbf{e}^{J}_b\delta_{IJ}\end{align}
 and the LSZ generalization reads, as first presented in \cite{Cheung:2021yog} (unbeknownst to us when derived in this work)
\begin{align}
    \mathcal S_{(n)}^{A_1\dots A_n}=&\frac{1}{Z[0]}\left(\prod_i^n(p_i^2-m_i^2)\mathbf{e}^{A_i}_{a_i}\frac{\de}{i\de J(p_i)_{a_i}}\right)Z[J]\\
    =&\left(\prod_i(p_i^2-m_i^2)\mathbf{e}^{A_i}_{a_i}\right)\mathcal G_{(n)}^{a_1\dots a_n}
\end{align}
a formula which can be verified readily with available results \cite{Alonso:2015fsp,Cohen:2021ucp} and reconciles the tensorial nature of matrix elements and amplitudes with their dependence on invariant geometric measures, as, e.g. in \cite{Cohen:2021ucp}, one can take a trace over the two to two $\mathcal S$ matrix to obtain a term proportional to the Ricci scalar.

\subsection{Quantum corrections}

After the definition of our covariant magnitudes, we turn to the evaluation of the partition function in an $\hbar$ expansion. Quantum corrections will be computed as Gaussian path integrals in Euclidean then rotated to Minkowski space; given our frequents trips between the two it is useful to make this explicit (the $(+---)$ metric being used):
\begin{align}
	x^0&=-ix^0_E\,, & p_0&=i(p_0)_E\,.
\end{align}
 We begin by shifting our integration from $V$ to a point $B$ for background on the manifold, given in RNC by $\eta_0$, and in $\phi$ coordinates by $\phi_0=\phi_{[V]}(\eta_0)$. It is around this point that we will perform a covariant expansion by deploying once more RNC, here denoted $\hat\eta_{\rm q}$. The relation between RNC around $V$, $\eta$, ($\eta\equiv\eta_0+\eta_{\rm q}$) and RNC coordinates around $B$, $\hat\eta_{\rm q}$, follows from both referring to the same point on the manifold, $\phi_{[V]}(\eta_0+\eta_q)=\phi_{[V]}(\eta_0)+\phi_{[B]}(\hat\eta_q)$, with the $\phi_{[B]}$ the mapping of eq.~(\ref{mapping}) around $B$. This relation can be cast as
 \begin{align}\label{Ann}
     \hat \eta_q^a&=(\phi_{[B]}^{-1})^a(\phi_{[V]}(\eta_0+\eta_q)-\phi_{[V]}(\eta_0))\\ \label{Ann2}
     &=\left(\frac{\partial\phi_{[V]}^a}{\partial\eta^i}\right)_{B} \eta_q^i\\\nonumber&+\frac12\left(\frac{\partial^2\phi_{[V]}^a}{\partial\eta^l\partial\eta^k}+\Gamma^{a}_{ij}\frac{\partial\phi_{[V]}^i}{\partial\eta^k}\frac{\partial\phi_{[V]}^j}{\partial\eta^l}\right)_{B}\eta_q^k\eta_q^l+\mathcal O(\eta_q^3)
 \end{align}
 and the source term in the action:
 \begin{align}\label{JdotEta}
     J\cdot \eta&=J\cdot \eta_0+J\cdot \eta_{\rm q}= J\cdot \eta_0+\hat J\cdot \hat\eta_{\rm q}+\mathcal O(\hat J \hat\eta_q^2)\,,\\
     \hat J_a&\equiv\left.\frac{\partial \eta^b}{\partial \phi^a_{[V]}}\right|_{B}J_b\,.\label{ParllT}
 \end{align}
 Relation (\ref{Ann2}) turns linear and is equivalent to parallel transport for $\eta_q\parallel\eta_0$ since all three $V$, $B$ and $\phi_V(\eta_0+\eta_q)$ points line up along a single geodesic. The limit $\eta_0\to 0$ gives the identity mapping $\hat \eta_q=\eta_q$ given $B\to V$ for $\eta_0\to0$ and hence $\phi_{[V]}(\eta_q)=\phi_{[V]}(\eta_q)$.  
 
The expansion in $\hat\eta_{\rm q}$ of our action is covariant as can be derived explicitly for the first few terms
\begin{align}\nonumber
\hat J_a&\hat\eta^a_{\rm q}+\left(\hat\eta^a_{\rm q}-\frac{\Gphi^a_{bc}}{2}\ \hat\eta^b_{\rm q}\hat\eta^c_{\rm q}\right)\frac{\de S}{\de\phi^a}\Bigg|_{B}+\frac12\hat\eta^2_{\rm q}\cdot\frac{\de^2S}{\de\phi^2}\Bigg|_{B}\!+\mathcal O(\hat\eta_{\rm q}^2\hat J)\\
=&\hat\eta^a_{\rm q}((\fD S[\phi_0])_a+\hat J_a)+\frac12\hat\eta^a_{\rm q}\hat\eta^b_{\rm q}(\fD^2\!S[\phi_0])_{ab}+\dots\,,
\end{align}
with $\fD$ the covariant functional derivative around $\phi_0$ which coincides to first order with the variation wrt $\phi$. As in the flat manifold case we define the point around which we expand by requiring the cancellation of the first term
 \begin{align}\label{phi0Def}
 &\frac{\de S}{\de\eta^a}[\eta_0]+J_a=\left.\frac{\partial\phi_{[V]}^b}{\partial\eta^a}\right|_{B}\fD_b S[\phi_0]+J_a\equiv0\\
 \label{phi0Def2}
 &\fD_a S[\phi_0]+\hat J_a\equiv 0\,,
 \end{align}
 where variation wrt $\eta_0$ (i.e. along $\eta_{\rm q}$) differs from variations wrt to $\hat\eta_{\rm q}$ as given by the chain rule (note that this holds regardless of neglecting higher orders in eq.~\ref{Ann2}) and the two definitions of $\eta_0$ being compatible as follows from eq.~(\ref{ParllT}). One has after this condition that all explicit $\hat J$ dependence is in higher order terms in eq.~(\ref{JdotEta}), $\mathcal O (\hat\eta^{n\geq 2}\hat J)=\mathcal O (-\hat\eta^{n\geq 2}\fD S)$. All these terms will not affect the $S$-matrix and further, are not necessary to obtain a covariant expansion of the action around $B$. It is for these reasons that we neglect them in the following, deferring their study and possible connection with redundancies in the partition function definition to future study. In practice this means using a linearised version of eq.~(\ref{JdotEta})
\begin{align}
     J\cdot \eta&=J\cdot \eta_0+J\cdot \eta_{\rm q}\to J\cdot \eta_0+\hat J\cdot \hat\eta_{\rm q}
 \end{align}
 the path integral reads
\begin{align}
Z[J]&=e^{iJ\cdot\eta_0}\int [d\hat\eta_{\rm q}]\sqrt{\det{\underline G}}\,e^{iS[\phi_0+\phi_{[B]}(\hat\eta_{\rm q})]+i\hat J\cdot\hat\eta_q}\,,
\end{align}
with $\phi(\hat\eta_q)$ as $\phi(\eta)$ in eq.~(\ref{mapping}) but with connections evaluated at $\phi_0$ rather than $0$.

The integral measure in RNC has itself a covariant expansion since the metric has the following expansion,
\begin{align}
\underline G_{ab}=&\left(\frac{\partial \phi}{\partial \eta}G(\phi_0+\phi_{[B]}(\hat\eta_{\rm q}))\frac{\partial \phi}{\partial \eta} \right)_{ab}\\=&G(\phi_0)_{ab}+\frac13\hat\eta^c_{\rm q}\hat\eta^d_{\rm q} R_{acdb}+\mathcal O(\eta^3_{\rm q})\,,
\end{align}
see \cite{Hatzinikitas:2000xe} for higher order terms.
The partition function with this expansion  is
  \begin{align}\nonumber
  	Z&=\int [d\hat\eta_{\rm q}] e^{\frac12{\rm Tr}\log(\underline G)+iS+iJ\cdot\eta_0+i\hat J\cdot\hat\eta_{\rm q}}\\
  	=& e^{iS[\phi_0]+iJ\cdot\eta_0+\frac{1}{2}{\rm Tr}\log(G(\phi_0))}\\ \nonumber
  	\times\!&\int [d\hat\eta_{\rm q}] {\rm Exp}\left( \sum_n\frac{\hat\eta^n_{\rm q}}{n!}\left[i\fD ^n S+\frac12\frac{\de^n}{\de\hat\eta^n_q}{\rm Tr}\log\left(\frac{\underline G}{G}\right)\right]_{B}\right).
  \end{align}
  Let us look at the second term in the measure series for $n=2$
  \begin{align}&
  \frac{1}{2!}\frac{1}{2}\hat\eta^2_{\rm q}\left[\frac{\de^2}{\de\hat\eta^2_{\rm q}}{\rm tr}\log\left(\frac{\underline G}{G}\right)\right]_{B}\\ \nonumber
  =&
  \frac{1}{12}\int \frac{d^dxd^dq}{(2\pi)^d}(G^{-1})^{cd}\hat\eta_{\rm q}^a\hat\eta_{\rm q}^b R_{acbd}  \propto \int d\Omega_{d}q^{d-2}dq^2\,.
  \end{align}
  Here one can observe both that the contribution is one loop order higher than $\fD^n S$ and that it cancels out if we use a regularization scheme without dimension-full parameters, e.g. dimensional regularization. This simplification will be applied later, but for the sake of generality here let us define
  \begin{align}\label{ovLS}
  	\overline{\fD^nS}=\left(\fD^n S-\frac{i}{2}\frac{\de^n}{\de \hat\eta^n_q}{\rm Tr}\log\left(\frac{\underline G}{ G}\right)\right)_{B}\,,
  \end{align}
  which makes the expansion,
  \begin{align}
  	Z[J]=& \mathcal N e^{i(S[\phi_0]+J\cdot \eta_0)-\frac{1}{2}{\rm Tr}[ \log(-(\overline{\fD^2S})G^{-1})]}\\ \nonumber
  	&\times \int[d\hat\eta_{\rm q}]\frac{\sqrt{\det(-\overline{\fD^2S})}}{\mathcal N}{\rm Exp}\left( \sum_{n=2}\frac i{n!}\hat\eta^n_{\rm q}\overline{\fD ^n S}\right),
  \end{align}
  where we divided and multiplied by the factor that results from the Gaussian integration.
  One last step defines
  \begin{align}
  \alpha\equiv \sqrt{-\overline{\fD^2 S}}\,\hat\eta_{\rm q}\,,
  \end{align} 
  where the square root operator has the  transformation properties of a vierbein, taking us from curved to flat field-space.
  This substitution gives us, now {\bf restoring} 
   $\hbar$ in our equations, (cf. eq.~(\ref{Bgng}) this is achieved by  $S\to S/\hbar, J\to J/\hbar$)
  \begin{align}
  Z[J]=& \mathcal N e^{\frac{i}{\hbar}(S[\phi_0]+J\cdot \eta_0)-\frac{1}{2}{\rm Tr}[ \log(-(\overline{\fD^2S})G^{-1})]}\\ \nonumber
  &\times\int [d\alpha]\, {\rm Gs}[\alpha]\,{\rm Exp}\left(i\sum_{n\geq3} \frac{\hbar^{n/2-1}\alpha^{n}\overline{\fD^{n}S}}{n!(-\overline{\fD^2 S})^{n/2}}\right)
  \end{align}
  with 
  \begin{align}
  {\rm Gs}[\alpha]\equiv & \frac{e^{\frac{i}{2}\alpha\cdot\alpha}}{\mathcal N},\
  &
  ({\rm Euclidean})\,\,\int[d\alpha]\,& {\rm Gs}[\alpha] =1\,.
  \end{align}
  We have arrived at our manifestly covariant loop expansion; for the one loop correction making indexes explicit,  $\overline{\fD^2S}G^{-1}=(\overline{\fD^2S})_a^{\,\,b}$, one can see the trace will produce an invariant term, whereas for the higher-order terms, indices of the $\overline{\fD^nS}$ tensors are taken from curved to flat field space by the verbein-like square-root of $-\overline{\fD^2S}$, where the functional $\alpha$ integrals are performed for an invariant result.
  
  While invariance is thus recovered, the loop expansion is somewhat obscured by the addition to $S$ that the measure causes in $\overline{ \fD^nS}$, eq.~(\ref{ovLS}), since in restoring $\hbar$ the two terms in $\overline{ \fD^nS}$ have different loop order. Were one to use dimensional regularization, this term would vanish leaving a simpler result, so let us write
  \begin{align}
  Z[J]=& \mathcal N Z_{\rm reg} e^{\frac i\hbar(S[\phi_0]+J\cdot \eta_0)-\frac{1}{2}{\rm Tr}[ \log(-(\fD^2S)G^{-1})]}\\ \nonumber
  &\times\int [d\alpha]\, {\rm Gs}[\alpha]\,{\rm Exp}\left(i\sum_{n\geq3} \frac{\hbar^{n/2-1}\alpha^{n}\mathcal  \fD^{n} S}{n!(-\fD^2 S)^{n/2}}\right)\,,
  \end{align}
  with $Z_{\rm reg}=1$ for dimensional regularization for any manifold or for a flat manifold in any regularization.
  {\it In the following we assume $Z_{\textrm{reg}}=1$}.

First let us make explicit that the operator indexes are contracted as
\begin{align}\nonumber
	\frac{\alpha^n \fD^n S}{(-\fD^2S)^{n/2}}&=(\fD^nS)_{\underline{x}_1...\underline{x}_n}\prod \left([ (-\fD^2 S)^{-1/2}]^{\underline{x}_i}_{\,\,\,\,\underline{y}_i}\alpha^{\underline{y}_i}\right) \\
	{\underline{x}}&=\{ x^\mu\,,\,a\}\,,
\end{align}
with our convention of integration/summation over repeated indexes and $\underline{x}$ capturing both space-time ($x^\mu$) and internal ($a$) indexes.
The n-th order loop correction to the generating functional we define as
\begin{align}\label{QnDef}
&\sum_{n\geq1} \hbar^n \mathcal{Q}_{n+1}\\ \nonumber
\equiv&\log\left(\int[d\alpha]\, \textrm{Gs}[\alpha]\,{\rm Exp}\!\left(\sum_{n\geq3} \frac{i\hbar^{n/2-1}\alpha^{n}\fD^{n}S}{n!(-\fD^2 S)^{n/2}}\right)\right)\,.
\end{align}
In the perturbative $\hbar$ expansion, the Taylor series of the exponential will yield integrals of Gaussian times polynomials which can be computed  by integrating by parts (in Euclidean)
\begin{align*}
\int [d\alpha]\, {\rm Gs}[\alpha] \prod_i^{2p}\alpha^{\underline x_i}=
\delta^{\underline{x_1x_2}}\dots\delta^{\underline{x_{p-1}x_p}}+{\rm ~perm.},
\end{align*} 
with $(2p-1)!!$ the number of terms, i.e. the number of possible pairings of $p$ elements in $p/2$ groups of 2, and $\de^{\underline{xy}}=\de^d(x-y)\de^{ab}$. Note that integration over an odd polynomial in $\alpha$ vanishes, this being the reason one need not compute $\sqrt{-\fD^2S}$ explicitly.

The operators obtained in this way are evaluated at the point $B$ of the manifold, that is, are functions of the field $\phi_0$ or equivalently the RNC $\eta_0$. The field itself is given in terms of the source $J$ by virtue of eq.~(\ref{phi0Def}) which allows the determination of the  $W[J]$
\begin{align}W[J]=&S[\phi_0(J)]+J\cdot\eta_0(J)+\sum\hbar^{n}\mathcal Q_n[\phi_0(J)]\,,
 \end{align}
  where we have included the one loop correction as $\mathcal{Q}_1$ and $\phi_0=\phi_{[V]}(\eta_0)$.
 The effective action then follows as
 \begin{align}\label{EffAi}
	\Gamma[\phi_{[V]}(\xi^V)]=&S[\phi_0]+J\cdot(\eta_0-\xi^V)+\sum \hbar^{n}\mathcal Q_n[\phi_0]\,,
\end{align}
where the effective field in RNC $\xi^V$ is given by the Legendre transform as
\begin{align}\nonumber
\xi^V\equiv&\frac{\de W}{\de J}=\frac{\de}{\de J}\left(S[\phi_{[V]}(\eta_0(J))]+J\cdot\eta_0+\sum_{n} \hbar^{n}\mathcal Q_n\right) \\
&=\eta_0+\frac{\de}{\de J}\left[ \sum \hbar^{n} \mathcal Q_n\right]\,,
\end{align}
where the cancellation of the term linear in $J$ follows from eq.~(\ref{phi0Def}) and to express the effective action as a explicit function of $\xi^V$ one should find its relation to $\eta_0$ and substitute it in. The field $\xi^V$ takes us to a third point in the manifold, let us call it $E$ and note that it is specified in $\phi$ coordinates by $\varphi\equiv\phi_{[V]}(\xi^V)$;
both $\xi^V$ and $\varphi$ are here referred to as effective fields. Points $B$ and $E$ are distinct, which is to say the background field $\phi_0$ differs from the effective field $\varphi$; it is useful to identify this difference via
\begin{align}
\Delta\xi^V&\equiv\eta_0-\xi^V=-\frac{\de}{\de J}\left[ \sum \hbar^{n} \mathcal Q_n\right]\,.\label{xiv}
\end{align}
which is related to RNCs around $B$, $\Delta\hat\xi$, by $\phi_{[V]}(\eta_0-\Delta \xi^V)=\phi_0+\phi_{[B]}(-\Delta\hat\xi)$, so that at the linear level
\begin{align}
\Delta\hat \xi=\left.\frac{\partial\phi_{[V]}}{\partial\eta}\right|_{B}\Delta \xi^V+\mathcal O(\Delta\xi^2)
\end{align}
\begin{figure}
\begin{tikzpicture}
\draw [thick] (-3,4) ..controls(0,3).. (3,4);
\draw [thick] (-3,3) ..controls(-1,2.4) and (1,2.2).. (3,3);
\draw [thick] (-3,2.25) .. controls (-0.5,1.65) and (2.5,1.5).. (3,1.8);
\draw [thick] (-3,1.2) .. controls (-1,1.1) and (2,1.2).. (3,0.8);
\draw [thick] (-3,0) ..controls(-1,0.5) and (2,1).. (3,-0.5);
\draw [thick] (-2,4) .. controls (-1.25,3) and (-1.5,2).. (-2,-0.5);
\draw [thick] (-1,4) .. controls (-0,2).. (-1,-0.5);
\draw [thick] (0.5,4) .. controls (1,2.5) and (1.3,1).. (0.3,-0.5);
\draw [thick] (2.3,4) .. controls (2.4,3) and (1.8,1).. (1.5,-0.5);
\draw [thick,purple] (-1.75,3.6) .. controls (1,2) .. (2,1.65);
\draw [thick,purple] (2,1.65) .. controls (1.5,1) and (0,0).. (-0.65,0.43);
\draw [->,thick] (-1.75,3.6)--(-1,3.05) node [anchor=north] {$\eta_0$};
\draw [->,thick] (-1.75,3.6)--(-1.15,4) node [anchor=south] {$\eta_{\rm q}$};
\draw [fill=blue] (-1.75,3.6) circle (2pt) node [anchor=north east] {$V$};
\draw [->,thick] (2,1.65)--(2.3,2.3) node [anchor=south west] {$\hat\eta_{\rm q}$};
\draw [->,thick] (2,1.65)--(1.45,0.9) node [anchor=north] {$-\Delta\hat\xi$};
\draw [fill=red] (2,1.65) circle (2pt) node [anchor=north west] {$B,\phi_0$};
\draw [->,thick] (-0.65,0.43)--(-0.1,0.1) node [anchor=north] {$\Delta\xi$};
\draw [fill=ForestGreen] (-0.65,0.43) circle (2pt) node [anchor=south east] {$E,\varphi$};
\end{tikzpicture}
	\caption{\label{Fig1} Scalar manifold parametrised by $\phi$ coordinates (black grid) with the three relevant points: vacuum $V$, background $B$, and effective field $E$ and RNCs at each point shown as tangent vectors.}
\end{figure}
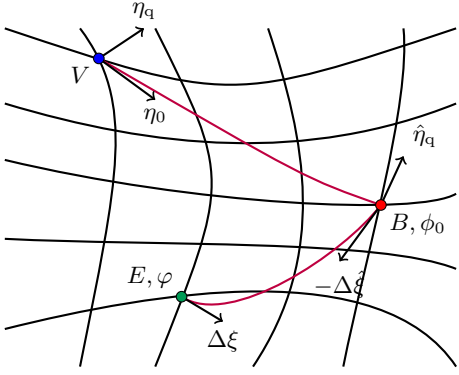
A relation which we will once more truncate to linear order and hence substitute
\begin{align}
    J\cdot (\eta_0-\xi^V)=J\cdot \Delta\xi^V
 \to\hat J\cdot \Delta\hat\xi  
\end{align}
As for the third manifold point, the conversion of $\Delta \hat\xi$ to RNC around $E$ is linear without any approximation, since $\Delta \hat \xi$ is tangent to the geodesic connecting $B$ and $E$ and connects the two points as $\varphi=\phi_0+\phi_{[B]}(-\Delta\hat\xi)$; so we  have
\begin{align}\label{DeDehat}
    \Delta \xi=\left(\frac{\partial \phi_{[B]} }{\partial\hat\eta}\right)_{E}\Delta \hat\xi
\end{align}
in terms of which $\varphi+\phi_{[E]}(\Delta\xi)=\phi_0$.
Manipulation of the tree-level action and source term yields, using our truncation and the operator $e^{\Delta\xi\fD}$ to shift back and forth in geodesic coordinates around $B$ and $E$
 \begin{align}
 &S[\phi_0]+\hat J\cdot \Delta\hat\xi  =S[\phi_0]-\Delta\hat\xi\fD S[\phi_0]\\ \nonumber
 =&(1-\Delta\hat\xi\fD)e^{\Delta\xi\fD}S[\varphi]=\left(1-\Delta\hat\xi\frac{\de}{\de\hat\eta}\right)e^{\Delta\xi\fD}S[\varphi]\\
 =&\left(1-\Delta\hat\xi\frac{\partial\phi_{[B]}}{\partial\hat\eta}\frac{\de}{\de \varphi}\right)e^{\Delta\xi\fD}S|_E=(1-\Delta\xi\fD)e^{\Delta\xi\fD}S[\varphi]\,,\nonumber
  \end{align} 
where the exponential is to be expanded in its Taylor series with derivatives evaluated at $E$. To make explicit the action of $\Delta\hat\xi\fD$ around $E$ rather than $B$ we used: the form $\Delta\hat\xi\fD=\Delta\hat\xi\delta/\delta\hat \eta$ with $\hat\eta$ RNCs around $B$, the chain rule, and the linear relation between $\Delta\hat\xi$ and $\Delta\xi$.
 The difference between background and effective field $\Delta \xi$ is found via eq.~(\ref{xiv}), first taking $\Delta\hat\xi$ and using the chain rule and eq.~(\ref{phi0Def}) to substitute $\fD \hat J=(-\fD^2S)^{-1}|_B$, finding 
\begin{align}\nonumber
\Delta \hat \xi&=(\fD^2S[\phi_0])^{-1}\fD  \sum_n\hbar^n \mathcal Q_n[\phi_0]
\end{align}
which can then be used to find $\Delta \xi$ via eq.~(\ref{DeDehat}) and has an expansion around $E$ as
\begin{align}\nonumber
\Delta\xi=&\left(\frac{\partial\phi_{[B]}}{\partial\hat\eta}\right|_E\left[\left(\frac{\partial\phi_{[B]}}{\partial\hat\eta}\right|_E\fD^2e^{\Delta \xi \fD}S[\varphi]\left(\frac{\partial\phi_{[B]}}{\partial\hat\eta}\right|_E \right]^{-1}\\&\times\left(\frac{\partial\phi_{[B]}}{\partial\hat\eta}\right|_E\fD e^{\Delta \xi \fD}Q[\varphi] \label{dephiEq}\\
=&(\fD^2 e^{\Delta\xi\fD}S[\varphi])^{-1}\fD\sum_n\hbar^n e^{\Delta\xi\fD}\mathcal Q_n[\varphi]\,,\label{dephiDef}
\end{align}
This definition has a recursive nature which nonetheless can be dealt with systematically in our $\hbar$ expansion,
\begin{align}
\Delta\xi=\sum_j \hbar^j\Delta\xi_j
\end{align}
with the two first terms as
\begin{align}\label{dephi1}
\Delta\xi_1^{\underline{x}}=&  [(\fD^2 S)^{-1}]^{\underline{xy}}\fD_{\underline{y}}\mathcal Q_1\,,\\
\Delta\xi_2^{\underline{x}}=& [(\fD^2 S)^{-1}]^{\underline{xy}}\fD_{\underline{y}}\mathcal Q_2\\ \nonumber&-[(\fD^2S)^{-1}]^{\underline{xy}}(\fD^2_{\underline{yz}} \Delta\xi_1\fD S)[(\fD^2S)^{-1}]^{\underline{zu}}\fD_{\underline{u}}\mathcal Q_1\\&+[(\fD^2S)^{-1}]^{\underline{xy}}(\fD_y \Delta\xi_1\fD \mathcal Q_1)\,.
\end{align}

Laborious as this might seem, the job of the Legendre transform is simply to get rid of the one-particle-reducible terms as usual and we shall show this explicitly in the next section.

To summarise, the functional method procedure for the effective action to order $\hbar^n$ requires
\begin{itemize}
	\item[\it i)]\label{LsT} computation, in Euclidean space, of Gaussian times polynomial path integrals to obtain $\mathcal Q_1$-$\mathcal Q_n$ from (\ref{QnDef}),
	\item[\it ii)] solving for the geodesic distance between the background $\phi_0$ and the effective field $\varphi$  (\ref{dephiDef},\ref{dephiEq}) to order $n-1$ $(\Delta\xi_{n-1})$,
\item [\it iii)] combining these results for the order $n$ effective action in
\begin{align}\label{CR}
\Gamma[\varphi]=(1-\Delta\xi\fD)e^{\Delta\xi\fD}S[\varphi]+e^{\Delta\xi\fD}\sum_{n} \hbar^{n} \mathcal Q_n[\varphi]\,.
\end{align}
\end{itemize}
This self contained procedure presents the loop expansion in QFT for any field manifold, with no restriction on the number of field insertions, and to any loop order. It comprises, just as in the case of CW \cite{Coleman:1973jx}, infinitely many Feynman diagrams whose summation beyond one loop with conventional methods might be unviable.

From this invariant effective action covariant correlation functions can be built, following standard methods, as 
\begin{align}
\mathcal G_n&
= \left(\Delta\xi+i\left(\fD^2 \Gamma\right)^{-1}\fD\right)^n\,,
\end{align}
\begin{align}\nonumber
\mathcal G_2&=i\left((\fD \Gamma)^{-1}\right|_{V},\qquad \mathcal G_3=i^2\left((\fD^2\Gamma)^{-3}\fD^3\Gamma\right|_{V},
\end{align}
which is the usual expression with $\fD_{\underline x}\Delta\xi^{\underline{y}}=\de^{\underline x}_{\underline y}$ since in geodesic coordinates the connection vanishes.
On passing we note that the correction from $\hbar\Delta\xi_1$ enters at the quadratic level on the $S[\varphi]$ expansion and multiplying a fist order quantum term $\mathcal Q_1$; it is for this reason that one can take $\phi_0=\varphi$ for the one loop result.



Lastly, from our definition of quantum corrections eq.~(\ref{QnDef}), a sum rule follows. Define
\begin{align}\nonumber
L=&~ \hbar{\rm~order}\,, & V&= ~\fD^{n>2}S {\rm ~insertions}\,, \\
I=&~ {\rm powers~of}~(\fD^2S)^{-1}\,, &V^{(j)}&=\fD^jS{\rm~insertions}\,,
\end{align}
with $V=\sum V^{(j)}$, one has the relation
\begin{align}
L-1&= I-V\,,& 
j \times V^{(j)}=2I\,.&
\end{align}
In diagrammatic terms, we would have the equivalence: $I$ as internal lines, $V$ as vertexes and $L$ as loops; this connection will be  made clearer in sec.~\ref{ThreeL} yet let us remark that no Feynman-diagram-like structure need be introduced in this formalism.


\subsection{The inverse of the second covariant variation of the action to order $\partial^2$}\label{S2}

To finalize the discussion in this section, let us derive an explicit inversion of $\fD^2S$ as this generalized `propagator'  is a building block of the expansion. For this purpose we restrict to an action at most quadratic in derivatives. While this is not needed for the formal derivation of results and it unavoidably somewhat reduces the generality, in practical examples an explicit form is the first element needed.

 Assuming asymptotic states with relativistic dispersion relations, let us take eq.~(\ref{KE}) as the leading action term with the addition of gauge interactions, gravity and a potential term as
\begin{align}
S=&\int d^dx\sqrt{|g|}\left[\frac12 {\rm d}_\mu \phi^a G_{ab}(\phi){\rm d}^{\mu}\phi^b-V_I\right]\\
	{\rm d}_\mu\phi&\equiv\partial_\mu\phi+A_\mu^I t^a(\phi)_I
\end{align}
with $t^a_I$ the I'th killing vector of our gauge symmetry which satisfy 
\begin{align}
	&({\rm L}_{t_I}(G))_{ab}=t^c_I\frac{\partial }{\partial\phi^c}G_{ab}+\frac{\partial t^c_I}{\partial \phi^a}G_{cb}+G_{ac}\frac{\partial t^c_I}{\partial \phi^b}=0\,,\\
	&[{\rm L}_{t_J},{\rm L}_{t_I}]=\frac{\partial t_{[J}^a}{\partial \phi^c}\frac{\partial t_{I]}^c}{\partial \phi^b}+t_{[J}^c\frac{\partial^2t_{I]}^a}{\partial\phi^c\partial\phi^b}=f_{IK}^{\,\,\,K}\frac{\partial t_K^a}{\partial\phi^b}\,,
\end{align}
where L is the lie derivative which makes the group structure more transparent yet it need and will not otherwise feature in the following.
The first variation reads
\begin{align}
	\hat\eta_{\rm q}\fD S=\int d^4x\sqrt{|g|}\,\hat\eta^a_{\rm q}(-\nabla_\mu(G_{ab} {\rm d}^\mu\phi^b)-\mathcal D_a V_I)
\end{align}
where
\begin{align}\label{CD0}
	&\nabla_\mu G_{ab} {\rm d}^\nu\phi^b\\\nonumber
	=&\left(g_{\rho}^\nu\partial_\mu\de_a^c-g_{\rho}^\nu A_\mu^I\frac{\partial t^c_I}{\partial\phi^a}-g_{\rho}^\nu\Gphi^c_{ad}{\rm d}_\mu\phi^d+\delta^c_a\Gamma^\nu_{\mu\rho}\right){\rm d}^\rho\phi_c\\ 
	\nonumber =&G_{ab}\left(g_\rho^\nu\partial_\mu\de_c^b+g^\nu_\rho A_\mu^I\frac{\partial t^b_I}{\partial\phi^b}+g^\nu_\rho\Gphi^b_{cd}d^\mu\phi^d+\delta^c_a\Gamma^\nu_{\mu\rho}\right){\rm d}^\rho\phi^c\,,
\end{align}
is our field, gauge and space-time covariant derivative with $\Gamma^\mu_{\nu\rho}$ the space-time connection. The second order variation is, in Euclidean, 
\begin{align}\label{D2S}
(-\fD^2 S)_{\underline{xy}}= \sqrt{|g|}\,\de^d(x-y)\left[-G_{ab}(\nabla)^2+U_{ab}\right]
\end{align}
with~\cite{Alonso:2016oah},
\begin{align}
	U_{ab}=(\mathcal D^2 V)_{ab}+\mathcal R_{cabd} {\rm d}_\mu\phi^c {\rm d}^\mu\phi^d
\end{align}
where $\mathcal{R}$ is the field Riemann tensor built with the field metric $G$;
\begin{align}
    \mathcal R^a_{\,\,bcd}=\partial_c \Gphi^a_{db}+\Gphi^a_{ce}\Gphi^e_{db}-(c \leftrightarrow d)\,.
\end{align}To solve for the inverse one can use Heat-Kernel techniques or, as it is done here, take an ansatz based on the CDE procedure \cite{Gaillard:1985uh,Henning:2014wua} as
\begin{widetext}
	\begin{align}
(-\fD^2 S)_{\underline{xy}}[(-\fD^2S)^{-1}]^{\underline{yz}}&=\int\sqrt{|g|}
 \de^d(x-y)\left(-G_{ab}(y)\nabla^2+U_{ab}\right)e^{iq(y-z)} \CdT{yq}^{-1} \mathscr O^{bc}\frac{d^dyd^dq}{(2\pi)^d}\\
&=\int\sqrt{|g|}
\de^d(x-y)e^{iq(y-z)}\CdT{yq}^{-1}\CdT{yq}\left(-G_{ab}(y)(iq+\nabla)^2+U_{ab}\right) \CdT{yq}^{-1} \mathscr O^{bc}\frac{d^dyd^dq}{(2\pi)^d}\\\nonumber
&=\int\sqrt{|g|}
e^{iq(x-z)}\CdT{xq}^{-1}\left[-G_{ab}\left(iq+\nabla+[i\partial_q,iq]\nabla+\mathcal O(1/q)\right)^2+ (U_{\CdT{}})_{ab}\right]\mathscr O^{bc}\frac{d^dq}{(2\pi)^d}\\
&\equiv\int\sqrt{|g|} e^{iq(x-z)}\CdT{xq}^{-1}\left[G_{ab}\left(q+\mathcal K([\nabla,\nabla],q)\right)^2+(U_{\mathcal{T}})_{ab}\right]\mathscr O^{bc}\frac{d^dq}{(2\pi)^d}
\end{align}
\end{widetext}
with $U_{\mathcal T}\equiv\CdT{yq} U\CdT{yq}^{-1}$ and the definition of our CDE transformation $\CdT{yq}$ in the last line is to remove `open' derivatives $\nabla$ acting to the right in favour of the commutator $[\nabla,\nabla]$ which allows for the equation of the inverse turning from differential to algebraic. The transformation in the presence of gravity is presently known to finite order in an expansion on inverse powers of $q$ \cite{Alonso:2019mok}, the first few terms being
\begin{align}\nonumber
\CdT{xq}&={\rm Exp}\left[\frac i2\left\{\partial_q^\mu,\nabla_\mu\right\}+\frac i4\left\{\left[\partial_q\nabla,\partial_q^\nu\right],q_\nu\right\}+\mathcal{O}\left(q^{-1}\right)\right],\\\nonumber
\mathcal K_\mu&= \frac{1}{4}\left\{\partial_q^\nu,\left[\nabla_\nu,\nabla_\mu\right]\right\}+\frac{1}{12}R^\nu_{\,\,\rho\kappa\mu} \left\{\partial_q^\rho\partial_q^\kappa,q_\nu\right\}+\mathcal{O}(q^{-2}),\\
U_{\CdT{}}&=U+\left[\nabla_\mu,U\right]\partial_q^\mu+\mathcal{O}(q^{-2})\,,\label{CdTwGR}
\end{align}
with $\{,\}$ the anticommutator.
The transformation 
is known exactly in the case without gravity where one has~\cite{Gaillard:1985uh,Henning:2014wua}
\begin{align}\nonumber
(\CdT{xq})_{g_{\mu\nu}\to\eta_{\mu\nu}}&={\rm Exp}\left[i\partial_q^\mu\nabla_\mu\right],\\\nonumber
(\mathcal K_\mu)_{g_{\mu\nu}\to\eta_{\mu\nu}}&=\sum i^nc_n\underbrace{[\partial_q\nabla,[...,[\partial_q\nabla}_{n{\rm ~times}},[\nabla_\mu,\nabla_\nu]\partial_q],...,],\\
(U_{\mathcal T})_{g_{\mu\nu}\to\eta_{\mu\nu}}&=\sum\frac{i^n}{n!}\underbrace{[\partial_q\nabla,[...,[\partial_q\nabla}_{n{\rm ~times}},U],...,]\,.
\end{align}
with $c_n=(n+1)/(n+2)!$.
Given these expressions, solving for $\mathscr{O}$ one obtains the inverse in Euclidean as 
	\begin{align}\nonumber
&\left[\frac{-1}{\fD^2S}\right]^{\underline{yz}}\!=\int\frac{d^dq}{(2\pi)^d}e^{iq(y-z)}\CdT{yq}^{-1} \frac{G^{ac}}{\sqrt{|g|}}
\mathcal{C}(q,\phi_0(y))_c^{\,\,b}\CdT{yq}
\\
&[\mathcal{C}(q,\phi_0(y))]_a^{\,\,b}\equiv\left[\frac{1}{(q+\mathcal K)^2+U_{\CdT{}}}\right]_a^{\,\,b}\,,\label{InvD2Def}
\end{align}
where the following expansion is to be understood in this expression 
\begin{align}
&q^2+\left\{q,\mathcal{K}\right\}+U_{\CdT{}}+\mathcal{K}^2\equiv q^2+U+ \Delta U\,,
\\
\label{mPexp}
&\frac{1}{(q+\mathcal{K})^2+U_{\CdT{}}}=\left[\frac{1}{q^2+U}\left(-\Delta U\right)\right]^n\frac{1}{q^2+U}\,.
\end{align}
The transformed covariant derivative when acting on an upper index $a$ is
\begin{align}
	\mathcal{K}_\mu\eta^a=&\frac12\left(F_{\mu\nu}^I\mathcal{D}_bt^a_I \partial_q^\nu+ \mathcal R^{a}_{\,\,cbd} {\rm d}_\mu\phi^c {\rm d}_\nu\phi^d\partial_q^\nu\right)\eta^b\\&+\frac{1}{12}R^\nu_{\,\,\rho\kappa\mu} \left\{\partial_q^\rho\partial_q^\kappa,q_\nu\right\}\eta^a+\mathcal{O}(q^{-2})\,,\nonumber
\end{align}
with $R_{\mu\nu}=R^{\alpha}_{\,\,\mu\alpha\nu}$.
Lastly to reduce the length of our expressions we define
\begin{align}\label{DefmP}
\mP\equiv (-\fD^2S)^{-1}\,.
\end{align}


\section{Three loop results}\label{ThreeL}

The perturbative expansion allows for derivation of results to any given loop order, to illustrate the obtention of such results we present here the three loop corrections. 

All $\hbar^{1/2}$ terms are odd in $\alpha$ and vanish when integrating; one encounters two terms to order $\hbar$ in eq.~(\ref{QnDef}), let us define:
\begin{align*}
\Gamma&=\sum_n \hbar^n\Gamma^{(n)}\,,
&
\mathcal Q_2&=\mathcal Q_{2,f}+\mathcal Q_{2,s}\,,
\end{align*}
so that the first contribution reads, in Euclidean,
\begin{align}\nonumber
\mathcal Q_{2,f}=&\int [d\alpha]\, {\rm Gs}[\alpha]\,\frac{1}{4} \frac{\alpha^4\fD^4S}{(-\fD^2S)^{2}}
\\&=
 \frac{3!!}{4!}(\fD^4S)_{\underline{xyzu}} \mP^{\underline{xy}}\mP^{\underline{zu}}\,,\label{Fly}
\end{align}
with integration and summation over repeated indexes and the 4th variation $\fD^4S$ is understood to be symmetrised on its indexes.  It is this index notation that makes it useful to introduce diagrams for bookkeeping, so the above looks like
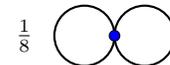
\begin{center}
	\begin{tikzpicture}
	\draw (1,0) node {$\frac{1}{8}$};
	\draw [thick] (1.8, 0) circle (0.4) ;
	\draw [thick] (2.6,-0) circle (0.4) ;
	\draw [fill=blue] (2.2,0) circle (2pt);
	\end{tikzpicture}
\end{center}
The index-exchange symmetry in $\fD^4S$ indexes yields a simple result.
For the second contribution some attention should be paid to the contraction of these indexes
\begin{align}\nonumber
\mathcal Q_{2,s}=&\int [d\alpha]\,{\rm Gs}[\alpha]\,\frac12\left(\frac{\alpha^3\fD^3S}{3! (-\fD^2S)^{3/2}}\right)^2\\ \nonumber=&
\frac{1}{8}\mP^{yz}(\fD^3S)_{\underline{xyz}}\mP^{\underline{xu}}(\fD^3S)_{\underline{uvw}}\mP^{\underline{vw}}\\&
+\frac{1}{12}(\fD^3S)_{\underline{xyz}}(\fD^3S)_{\underline{uvw}}\mP^{\underline{xu}}\mP^{\underline{yv}}\mP^{\underline{zw}}\label{SSt}\,.
\end{align}
These two terms can be represented diagrammatically as 
\begin{center}
	\begin{tikzpicture}
	\draw (0.8,0) node {$\frac{1}{8}$};
	\draw [thick] (1.5, 0) circle (0.4) ;
	\draw [thick] (1.9,0) -- (2.6,0);
	\draw [thick] (3,-0) circle (0.4) ;
	\draw [fill=blue] (1.9,0) circle (2pt);
	\draw [fill=blue] (2.6,0) circle (2pt);
	\draw (3.9,0) node {$+\frac{1}{12}$};
	\draw [thick] (5,0) circle (0.5);
	\draw [thick] (4.5,0) -- (5.5,0);	
	\draw [fill=blue] (4.5,0) circle (2pt);
	\draw [fill=blue] (5.5,0) circle (2pt);
	\end{tikzpicture}
\end{center}
The effective action at two loops requires of the definition of the effective (Legendre transformed) field to one loop, eq.~(\ref{dephiEq}), which yields a two loop correction in addition to $\mathcal{Q}_2$
\begin{align}
\Gamma^{(2)}[\varphi]=-\frac12(\Delta\xi_1)^2\cdot\fD^2S+\Delta\xi_1\cdot\fD\mathcal Q_1+\mathcal Q_2\,,
\end{align}
where from eq.~(\ref{dephi1}) follows that 
\begin{align}
(\Delta\xi_1)^{\underline{x}}=-
\frac{1}{2}\mP^{\underline{xy}}(\fD^3S)_{\underline{yzu}}\mP^{\underline{zu}}\,,
\end{align}
and one has
\begin{align}
&-\frac12(\Delta\xi_1)^2\fD^2S+\Delta\xi_1\fD\mathcal Q_1\\\nonumber&=-\frac18\mP^{\underline{yz}} (\fD^3S)_{\underline{xyz}}\mP^{\underline{xu}}(\fD^3S)_{\underline{uvw}}\mP^{\underline{vw}}\,.
\end{align}
Diagrammatically one can recognize
\begin{center}
	\begin{tikzpicture}
	\draw (0.6,0) node {$-\frac{1}{8}$};
	\draw [thick] (1.5, 0) circle (0.4) ;
	\draw [thick] (1.9,0) -- (2.6,0);
	\draw [thick] (3,-0) circle (0.4) ;
	\draw [fill=blue] (1.9,0) circle (2pt);
	\draw [fill=blue] (2.6,0) circle (2pt);
	\end{tikzpicture}
\end{center}
which will cancel out against the first term in the Gaussian integral in eq.~(\ref{SSt}). It is here that one sees the Legendre transform at work subtracting all one-particle-reducible contributions.

Turning back to to {\it Minkowski} the Effective action to the two loop order reads
\begin{align}\nonumber
\Gamma[\varphi]=&S[\varphi]-\frac i 2 \hbar\, {\rm Tr}\log(\mP)
{-}\frac{\hbar^2}{8}\fD^4S_{\underline{xyzw}}\mP^{\underline{xy}}\mP^{\underline{zw}}\\
	&{-}\frac{\hbar^2}{12} \fD^3S_{\underline{xyz}}\mP^{\underline{xw}}\mP^{\underline{yv}}\mP^{\underline{zu}} \fD^3S_{\underline{uvw}}+\mathcal O(\hbar^3)
\end{align}
Or, diagrammatically,
\begin{center}
	\begin{tikzpicture}
	\draw (-1.5,0) node {$-\frac{i}{2}$};
	\draw [thick] (0,0) circle (0.5);
	\draw (-0.85,0) node {log};
	\draw (1.5,0) node {$-\frac{1}{8}$};
	\draw [thick] (2.5, 0.4) circle (0.4) ;
	\draw [fill=blue] (2.5,0) circle (2pt);
	\draw [thick] (2.5,-0.4) circle (0.4) ;
	\draw (4,0) node {$-\frac{1}{12}$};
	\draw [thick] (5,0) circle (0.5);	
	\draw [fill=blue] (4.5,0) circle (2pt);
	\draw [fill=blue] (5.5,0) circle (2pt);
	\draw [thick] (4.5,0) -- (5.5,0);
	\end{tikzpicture}
\end{center}
with n-vertexes representing $\fD^n S$ and lines  $\mP=(-\fD^2S)^{-1}$, this being the reason we substitute $\mP$ inside the trace rather than the usual $(-\fD^2S)$ and hence the extra sign.

Two loop terms present all elements of the computation in simplified form; for the three loop order one has
\begin{align}\nonumber\
\Gamma^{(3)}[\varphi]=&-\frac13(\Delta\xi_1)^3\fD^3S-\Delta\xi_1\Delta\xi_2 \fD^2S+\frac12(\Delta\xi_1)^2\fD^2\mathcal Q_1\\&+\Delta\xi_2\fD \mathcal Q_1+\Delta\xi_1\fD\mathcal Q_2+\mathcal Q_3
\end{align}
where the first five terms give one particle reducible terms that cancel out against the terms in the Gaussian integrals of $\mathcal Q_3$; e.g. the term
\begin{align}
    -\frac{1}{48}\fD^3 S_{\underline{x_1x_2x_3}}\prod_i (\mP^{x_iy_3}\fD^3 S_{\underline{y_1^iy_2^iy_3^i}}\mP^{\underline{y_1^iy_2^i}}),
\end{align}
\begin{center}
	\begin{tikzpicture}
	\draw (-2,0) node {$-\frac{1}{48}$};
	\draw [thick] (0 , 1) circle (0.4) ;
	\draw [thick] (0.87 , -0.5) circle (0.4);
	\draw [thick] (-0.87,-0.5) circle (0.4) ;
	\draw [thick] (0,0) -- (0,0.6);
	\draw [thick] (0,0) -- (0.52,-0.3);
	\draw [thick] (0,0) -- (-0.52,-0.3);
	\draw [fill=blue] (0,0) circle (2pt);
	\draw [fill=blue] (-0.52,-0.3) circle (2pt);
	\draw [fill=blue] (0,0.6) circle (2pt);
	\draw [fill=blue] (0.52,-0.3) circle (2pt);
	\end{tikzpicture}
\end{center}
cancels against an opposite sign contribution from $\mathcal Q_3$, as do the other 6 one-particle reducible contributions. 
The outcome of the procedure gives the effective action to three loops, in Minkowski,
\begin{widetext} 
\begin{align}\nonumber
	\Gamma[\varphi]=&S[\varphi]-\frac{i\hbar}2[\log (\mP)]_{\underline x}^{\,\,\,\underline x}
	-\frac{\hbar^2}{8}\fD^4S_{\underline{xyzw}}\mP^{\underline{xy}}\mP^{\underline{zw}}
	-\frac{\hbar^2}{12} \fD^3S_{\underline{xyz}}\mP^{\underline{xw}}\mP^{\underline{yv}}\mP^{\underline{zu}} \fD^3S_{\underline{uvw}}
	\\ \nonumber
	&+\frac{i\hbar^3}{48} \fD^6S_{\underline{xyzuvw}}\mP^{\underline{xy}}\mP^{\underline{zu}}\mP^{\underline{vw}}+\frac{i\hbar^3}{12}\mP^{\underline{xy}}\fD^5S_{\underline{xyzuv}}\mP^{\underline{zw}}\mP^{\underline{us}}\mP^{\underline{vt}}\fD^3S_{\underline{wst}}\\ \nonumber
	&+\frac{i\hbar^3}{48}\fD^4S_{\underline{xyzu}}\mP^{\underline{xv}}\mP^{\underline{yw}}\mP^{\underline{zs}}\mP^{\underline{ut}}\fD^4S_{\underline{vwst}}+\frac{i\hbar^3}{16}\mP^{\underline{xy}}\fD^4\!S_{\underline{xyzu}}\mP^{\underline{zv}}\mP^{\underline{uw}}\fD^4S_{\underline{vwst}}\mP^{\underline{st}}\\  \nonumber&+\frac{i\hbar^3}{8}\fD^4S_{\underline{xyzu}}\mP^{\underline{xv}}\mP^{\underline{yw}}\fD^3S_{\underline{vws}}\mP^{\underline{zt}}\mP^{\underline{ur}}\fD^3S_{\underline{trj}}\mP^{\underline{js}}+\frac{i\hbar^3}{8}\mP^{\underline{xy}}\fD^4S_{\underline{xyzu}}\mP^{\underline{zv}}\fD^3S_{vsj}\mP^{\underline{uw}}\fD^3S_{\underline{wtr}}\mP^{\underline{st}}\mP^{\underline{jr}}\\\nonumber&+\frac{i\hbar^3}{16}\fD^3S_{xyz}\mP^{\underline{xu}}\mP^{\underline{ys}}\mP^{\underline{zt}}\fD^3S_{\underline{str}}\fD^3S_{uvw}\mP^{\underline{vi}}\mP^{\underline{wj}}\fD^3S_{ijm}\mP^{\underline{mr}}\\
	&+\frac{i\hbar^3}{8}\mP^{\underline{tx}}(\fD^3S)_{\underline{xyz}}\mP^{\underline{yu}}(\fD^3S)_{\underline{uvw}}\mP^{\underline{vs}}(\fD^3S)_{\underline{str}}\mP^{\underline{zi}}\mP^{\underline{wj}}\mP^{\underline{rm}}(\fD^3S)_{\underline{ijm}}+\mathcal{O}(\hbar^4)\label{3LG}
\end{align}
or diagrammatically, 
\begin{center}
	\begin{tikzpicture}
	\draw (-3.5,0) node {$\Gamma\,=$};
	\draw (-2.5,0) node {$S$};
	\draw (-1.4,0) node {$-\frac{i}{2}$};
	\draw [thick] (0,0) circle (0.5);
	\draw (-0.85,0) node {log};
	\draw (1.5,0) node {$-\frac{1}{8}$};
	\draw [thick] (2.5, 0.4) circle (0.4) ;
	\draw [fill=blue] (2.5,0) circle (2pt);
	\draw [thick] (2.5,-0.4) circle (0.4) ;
	\draw (4,0) node {$-\frac{1}{12}$};
	\draw [thick] (5,0) circle (0.5);	
	\draw [fill=blue] (4.5,0) circle (2pt);
	\draw [fill=blue] (5.5,0) circle (2pt);
	\draw [thick] (4.5,0) -- (5.5,0);
	\end{tikzpicture}
\raisebox{6pt}{	\begin{tikzpicture}
	\draw (-1.4,0) node {$+\frac{i}{8}$};
	\draw [thick] (-0.5, 0) circle (0.5) ;
	\draw [thick] (0.5,0.5) -- (0.5,-0.5);
	\draw [thick] (0.5,0) circle (0.5) ;
	\draw [fill=blue] (0,0) circle (2pt);
	\draw [fill=blue] (0.5,0.5) circle (2pt);
	\draw [fill=blue] (0.5,-0.5) circle (2pt);
	\end{tikzpicture}}\quad
\raisebox{9pt}{
	\begin{tikzpicture}
	\draw (-1.47,0) node {$+\frac{i}{12}$};
	\draw [thick] (-0.5, 0) circle (0.5) ;
	\draw [thick] (0,0) -- (1,0);
	\draw [thick] (0.5,0) circle (0.5) ;
	\draw [fill=blue] (0,0) circle (2pt);
	\draw [fill=blue] (1,0) circle (2pt);
	\end{tikzpicture}}\\\qquad
	\raisebox{3pt}{\begin{tikzpicture}
	\draw (-1.3,0) node {$+\frac{i}{16}$};
	\draw [thick] (-0.40, 0) circle (0.40) ;
	\draw [thick] (0.40,0) circle (0.40) ;
	\draw [thick] (1.2,0) circle (0.4) ;
	\draw [fill=blue] (0,0) circle (2pt);
	\draw [fill=blue] (.8,0) circle (2pt);
	\end{tikzpicture}}\quad	
	\begin{tikzpicture}
	\draw (-1,0) node {$+\frac{i}{8}$};
	\draw [thick] (0,0) circle (0.5);
	\draw [thick] (-0.5,0) arc (90:36:1);
	\draw [thick] (-0.5,0) arc (270:324:1);
	\draw [fill=blue] (-.5,0) circle (2pt);
	\draw [fill=blue] (0.3,0.38) circle (2pt);
	\draw [fill=blue] (0.3,-0.38) circle (2pt);
	\end{tikzpicture}\quad
	\begin{tikzpicture}
	\draw (-1.2,0) node {$+\frac{i}{16}$};
	\draw [thick] (0,0) circle (0.5);
	\draw [thick] (0,0.17) arc (270:295:1);
	\draw [thick] (0,0.17) arc (270:245:1);
	\draw [thick] (0,-0.17) arc (90:65:1);
	\draw [thick] (0,-0.17) arc (90:115:1);
	\draw [fill=blue] (0.41,0.26) circle (2pt);
	\draw [fill=blue] (0.41,-0.26) circle (2pt);
	\draw [fill=blue] (-0.41,0.26) circle (2pt);
	\draw [fill=blue] (-0.41,-0.26) circle (2pt);
	\end{tikzpicture}\quad
	\begin{tikzpicture}
	\draw (-1.1,0) node {$+\frac{i}{48}$};
	\draw [thick] (0,0) circle (0.5);
	\draw [thick] (-0.5,0) arc (240:300:1);
	\draw [thick] (0.5,0) arc (60:120:1);
	\draw [fill=blue] (0.5,0) circle (2pt);
	\draw [fill=blue] (-0.5,0) circle (2pt);
	\end{tikzpicture}\quad
	\raisebox{5pt}{\begin{tikzpicture}
	\draw (-1,0) node {$+\frac{i}{48}$};
	\draw [thick] (0,0.5) arc (120:240:0.305);
	\draw [thick] (0,0.5) arc (60:-60:0.305);
	\draw [thick] (0,0) arc (60:180:0.305);
	\draw [thick] (0,0) arc (0:-120:0.305);
	\draw [thick] (0,0) arc (120:0:0.305);
	\draw [thick] (0,0) arc (180:300:0.305);
	\draw [fill=blue] (0,0) circle (2pt);
	\end{tikzpicture}}
	\begin{tikzpicture}
	\draw (-1,0) node {$+\frac{i}{8}$};
	\draw [thick] (0,0) circle (0.5);
	\draw [thick] (0,0) -- (0,0.5);
	\draw [thick] (0,0) -- (0.43,-0.25);
	\draw [thick] (0,0) -- (-0.43,-0.25);
	\draw [fill=blue] (0,0) circle (2pt);
	\draw [fill=blue] (0,0.5) circle (2pt);
	\draw [fill=blue] (0.43,-0.25) circle (2pt);
	\draw [fill=blue] (-0.43,-0.25) circle (2pt);
	\end{tikzpicture}\raisebox{12pt}{\,\,+\,$\mathcal O(\hbar^4)$\,.}
\end{center}
where we note that the coefficient of each term can also be derived with symmetry factor formulae \cite{Palmer:2001vq,Saemann:2020oyz}. Let us remark that this formula applies to any tree-level action $S$, regardless of the number of derivatives or powers of fields in interaction terms.
 The explicit evaluation of the operators at two loops, for the particular case of an action as in sec.~\ref{S2} would involve the integrals, in Euclidean,
	\begin{align}\nonumber
	\fD^4S_{\underline{xyzw}}\mP^{\underline{xy}}\mP^{\underline{zw}} 
	=&\int \frac{dxdydzdudqdp}{(2\pi)^{2d}\sqrt{|g_x|}\sqrt{|g_z|}}(\fD^4S)_{\underline{xyzu}} e^{iq(x-y)}\CdT{xq}^{-1}\left[\mathcal C(q,\partial_q,x)\right]^{ab}\CdT{xq}e^{ip(z-u)}\CdT{zp}^{-1}\left[\mathcal C(p,\partial_p,z)\right]^{cd}\CdT{zp}\\
	=&\int \frac{dxdpdq}{(2\pi)^{2d}\sqrt{|g|}}[(\fD^4\!\mathcal L)(x)]_{abcd} \left(\CdT{xq}^{-1}\left[\mathcal C(q,\partial_q,x)\right]^{ab}\CdT{xq}\right)\left(\CdT{xp}^{-1}\left[\mathcal C(p,\partial_p,x)\right]^{cd}\CdT{xp}\right)\label{FlyFull}
	\end{align}
 where we assumed the locality of the interaction can be cast as $(\fD^4S)_{\underline{xyzu}}\equiv \sqrt{|g|}(\fD^4\mathcal L)_{abcd}\de^d(x-y)\de^d(x-z)\de^d(x-u)$ and we note that the momenta integrals will restore the invariant measure of space-time ($q^2=g^{\mu\nu} q_\mu q_\nu$). The second term at two loops requires of an expansion in the two vertexes separation $x-y$ to extract local results, explicitly
		\begin{align}\label{SunSetFull}
	&\fD^3S_{\underline{x_1x_2x_3}}\mP^{\underline{x_1y_1}}\mP^{\underline{x_2y_2}}\mP^{\underline{x_3y_3}} \fD^3S_{\underline{y_1y_2y_3}}\\\nonumber
		=&\int \left(\prod_{j=1}^3\frac{dx_jdy_j}{\sqrt{|g_{x_i}|}}\right) (\fD^3S)_{\underline{x_1x_2x_3}}\fD^3S_{\underline{y_1y_2y_3}}\left(\prod_{i=1}^3\frac{e^{i p_i(x_i-y_i)}dq_i}{(2\pi)^{d}}\CdT{p_ix_i}^{-1}\left[\mathcal C(p_i,\partial_{p_i},x_i)\right]^{a_ib_i}\CdT{p_ix_i}\right)
	\\
	=&\int \frac{dxdy}{|g_x|}[\fD^3 \mathcal L(x)]_{a_1a_2a_3}\sqrt{|g_y|}[\fD^3\mathcal L(y)]_{b_1b_2b_3}e^{i\sum p_i(x-y)}\left(\prod_{i=1}^3\frac{dq_i}{(2\pi)^{d}}\CdT{p_ix}^{-1}\left[\mathcal C(p_i,\partial_{p_i},x)\right]^{a_ib_i}\CdT{p_ix}\right)\nonumber\\ \nonumber
		=&\int \frac{dx}{|g_x|}[\fD^3\!\mathcal L(x)]_{a_1a_2a_3}\left(\prod_{i=1}^3\frac{dq_i}{(2\pi)^{d}}\CdT{p_ix_i}^{-1}\left[\mathcal C(p_i,\partial_{p_i},x_i)\right]^{a_ib_i}\CdT{p_ix_i}\right)e^{(y-x)\nabla_x}\sqrt{-g_x}[\fD^3\mathcal L(x)]_{b_1b_2b_3}e^{-(x-y)\nabla_x}e^{i\sum p_i(x-y)}dy\nonumber\\ 
		=&\int \frac{dx}{|g_x|}\,[\fD^3\!\mathcal L(x)]_{a_1a_2a_3}\left(\prod_{i=1}^3\frac{dq_i}{(2\pi)^{d}}\CdT{p_ix_i}^{-1}\left[\mathcal C(p_i,\partial_{p_i},x_i)\right]^{a_ib_i}\CdT{p_ix_i}\right) \left(\CdT{p_1x}\left[ \sqrt{|g_x|}[\fD^3\!\mathcal L(x)]_{{b_1b_2b_3}}\right] \CdT{p_1x}^{-1}(2\pi)^d\de(\sum p_i)\right)\,,\nonumber
	\end{align}
\end{widetext}
	where we have used  overall invariance of the expression to move from $x$ to $y$ with a covariant derivative expansion and derivatives act within the parenthesis. These expressions can expanded on field derivatives via eq.~(\ref{mPexp}) with the first order given the effective potential corrections, the second the kinetic term etc.



\section{Two loop effective potential for a N-manifold with O(N) symmetry}\label{TwoLVeff}

Consider the tree level action with an $\mathcal O(N)$ symmetry:
\begin{align}
	S=\int d^4x \frac12\left(\partial_\mu\phi^a G_{ab}\partial^\mu\phi^b-V(\phi)\right),
\end{align}
	where the index in $\phi^{a}$ runs over $N$ fields: a radial mode $h$ and $n=N-1$ angular modes $\varphi^{i=1,...,n}$ parametrizing a $S^{N-1}$ sphere where the $O(N)$ symmetry acts. The metric reads
\begin{align}
	G_{ab}&=\left(\begin{array}{cc}
	1&\\
	&F^2(h)\tilde g_{ij}
	\end{array}\right),
	&G^{ab}&=\left(\begin{array}{cc}
	1&\\
	&\tilde g^{ij}/F^2(h)
	\end{array}\right).
\end{align}
	with $\tilde{g}$ being the n-sphere metric, which can be given in terms of an $N$-dimensional unit vector $u(\varphi)$ as
\begin{align}
    \tilde g_{ij} = \frac{\partial u(\varphi)}{\partial \varphi^i}\cdot \frac{\partial u(\varphi)}{\partial \varphi^j}, & & u \cdot u=1.
\end{align}
Maximally symmetric cases would give a scalar field manifold  as R$^N$, S$^N$ or H$^N$  with explicit $F(h)$ functions for each yet it could also be other type of manifold, one with e.g. non-trivial topology; to capture all these and more we keep $F(h)$ general.

A number of tensors will feature in the two loop effective potential;  first, those related to the curvature of field space read,
\begin{align}
		\mathcal R_{hijh}&=-F^2\mathcal R_h\tilde g_{ij},\\
		\mathcal R_{ijkl}&= F^4\mathcal R_\varphi (\tilde g_{ik}\tilde g_{jl}-\tilde g_{il}\tilde g_{jk}),
\end{align}
in addition to entries related by the symmetry properties of the Riemann tensor and where,
\begin{align}\label{Rdef}
	\mathcal R_h&\equiv-\frac{F''}{F}, & \mathcal R_\varphi&\equiv\frac{1}{v^2F^2}-\frac{(F')^2}{F^2};
\end{align}
second, tensors built out of the potential at the second:
\begin{align}
	(\mathcal D^2V)_{ab}=\left(\begin{array}{cc}
	V''&\\
	&F'FV'\tilde g_{ij}
	\end{array}\right),
\end{align}
third
\begin{align}
	(\mathcal D^3V)_{hhh}&=V''',\\
	(\mathcal D^3V)_{hij}&=F^2\Big(\frac{V'F'}{F}\Big)'\tilde g_{ij},\\
	(\mathcal D^3V)_{ihj}&=(\mathcal D^3V)_{ijh}=F^2F'\Big(\frac{V'}{F}\Big)'\tilde g_{ij},
\end{align}
and fourth order
\begin{align}
	(\mathcal D^4V)_{hhhh}&=V'''',\\
	(\mathcal D^4V)_{hhij}&=F^2\Big(\frac{V'F'}{F}\Big)''\tilde g_{ij},\\
	(\mathcal D^4V)_{hihj}&=(\mathcal D^4V)_{hijh}=F^2\Big(F'\Big(\frac{V'}{F}\Big)'\Big)'\tilde g_{ij},\\
	(\mathcal D^4V)_{ihhj}&=(\mathcal D^4V)_{ihjh}\nonumber\\
	&=FF'\Big(FF'V'''-\Big(\frac{V'F'}{F}\Big)'-F'\Big(\frac{V'}{F}\Big)'\Big)\tilde g_{ij},\\
	(\mathcal D^4V)_{ijhh}&=FF'\Big(V'''-2F'\Big(\frac{V'}{F}\Big)'\Big)\tilde g_{ij},\\
	(\mathcal D^4V)_{ijkl}&=F^3F'\left(\Big(\frac{V'F'}{F}\Big)'\tilde g_{ij} \tilde g_{kl}+F'\Big(\frac{V'}{F}\Big)' \tilde g_{i(k} \tilde g_{l)j}\right).
\end{align}

These will feature in the second, third and fourth order covariant variations of the action that enter the two-loop computation. The second order variation has been given in eq.~(\ref{D2S}); with the added simplification that follows from neglecting gauge and gravitational interactions,
i.e. eq.~(\ref{CD0}) now reads  $\nabla_\mu \eta^a= \partial_\mu \eta^a - \Gamma^a_{bc}(\partial_\mu \phi)^b \eta^c$,
and restricting to the effective potential which further allows to set $\partial_\mu\phi\to 0$, the inverted second order variation $\mP$ reads,
 in terms of the coefficient $\mathcal C$ in eq.~(\ref{InvD2Def}), in Euclidean,
\begin{equation}
\begin{aligned}
	\left[\mathcal C_{\partial\phi\to 0}(q)\right]^a_b
	&=\left[\frac{1}{q^2+\mathcal D^2V}\right]_b^a\\
	&=\left(\begin{array}{cc}
	\frac{1}{(q^2+V'')}&\\&\frac{ \delta^{i}_j}{q^2+F'V'/F}
	\end{array}
	\right).
\end{aligned}
\end{equation}
Diagrammatically, we associate the above with a line. Next, the third and fourth order variations, which we associate with 3- and 4-point vertices respectively, have the form, in Minkowski:
\begin{equation}
\begin{aligned}
\hat\eta_{\rm q}^3\fD^3S&=\int d^4x\Big\{\mathcal D_{c_1} \mathcal R_{ac_2c_3b}\hat\eta_{\rm q}^{c_1}\hat\eta_{\rm q}^{c_2}\hat\eta_{\rm q}^{c_3}\partial_\mu \phi^a \partial^\mu\phi^b\\
&+4\mathcal R_{ac_1c_2c_3}\hat\eta_{\rm q}^{c_1}\hat\eta_{\rm q}^{c_2} \nabla_\mu \hat\eta_{\rm q}^{c_3} \partial^\mu \phi^a-\hat\eta_{\rm q}^a\hat\eta_{\rm q}^b\hat\eta_{\rm q}^c(\mathcal D^3  V)_{abc}\Big\},
\end{aligned}
\end{equation}
and \cite{Alvarez-Gaume:1981exa},
\begin{equation}
\begin{aligned}
	\hat\eta_{\rm q}^4\fD^4S&=\int d^4x\Big\{ 4\mathcal R_{c_1c_2c_3c_4}\hat\eta_{\rm q}^{c_2}\hat\eta_{\rm q}^{c_3}\nabla_\mu \hat\eta_{\rm q}^{c_1} \nabla^\mu \hat\eta_{\rm q}^{c_4}\\
	&+6\mathcal D_{c_1} \mathcal R_{ac_2c_3c_4}\hat\eta_{\rm q}^{c_1}\hat\eta_{\rm q}^{c_2}\hat\eta_{\rm q}^{c_3} \nabla_\mu \hat\eta_{\rm q}^{c_4} \partial^\mu \phi^a\\
	&+(\mathcal D_{c_1} \mathcal D_{c_2} \mathcal R_{ac_3c_4b} +4\mathcal R^d_{\,\,c_1c_2a}\mathcal R_{dc_3c_4b})\\
	&\qquad\qquad\hat\eta_{\rm q}^{c_1}\hat\eta_{\rm q}^{c_2}\hat\eta_{\rm q}^{c_3}\hat\eta_{\rm q}^{c_4}\partial_\mu \phi^a \partial^\mu \phi^b\\
	&-\hat\eta_{\rm q}^a\hat\eta_{\rm q}^b\hat\eta_{\rm q}^c\hat\eta_{\rm q}^d(\mathcal D^4V)_{abcd} \Big\},
\end{aligned}
\end{equation}
where we note that the exchange symmetry of the quantum fields $\eta$ selects the fully symmetric components of each tensor. This will give the symmetric covariant derivative tensor of the potential in each case, whereas for the curvature terms with covariant derivatives acting on $\eta$, this symmetrization would yield, in momentum space using our CDE expansion, in Minkowski,
\begin{align}
&4\mathcal R_{abcd}\nabla_\mu\hat\eta_{\rm q}^a\hat\eta_{\rm q}^b\hat\eta_{\rm q}^c\nabla^\mu\hat\eta_{\rm q}^d\nonumber\\
    =&\frac{1}{3}\Big(\mathcal R_{abcd} (iq_{(a)}-iq_{(b)})(iq_{(d)}-iq_{(c)})\nonumber\\
	&\quad+\mathcal R_{acbd} (iq_{(a)}-iq_{(c)})(iq_{(d)}-iq_{(b)})\\ \nonumber
	&\quad+\mathcal R_{adcb} (iq_{(a)}-iq_{(d)})(iq_{(b)}-iq_{(c)}) \Big)\hat\eta_{\rm q}^a\hat\eta_{\rm q}^b\hat\eta_{\rm q}^c\hat\eta_{\rm q}^d+\mathcal O (\mathcal K)
\end{align}

Let us separate the two diagrams which contribute at two-loop order to the effective potential as, 
\begin{align}
    -\int d^4x V^{(2)}_{\rm eff} &=\begin{tikzpicture}[baseline={([yshift=-.5ex]current bounding box.center)}]
	\draw (1.5,0) node {$-\frac{1}{8}$};
	\draw [thick] (2.5, 0.4) circle (0.4) ;
	\draw [fill=blue] (2.5,0) circle (2pt);
	\draw [thick] (2.5,-0.4) circle (0.4) ;
	\draw (3.5,0) node {$-\frac{1}{12}$};
	\draw [thick] (4.5,0) circle (0.5);	
	\draw [fill=blue] (4,0) circle (2pt);
	\draw [fill=blue] (5,0) circle (2pt);
	\draw [thick] (4,0) -- (5,0);
	\end{tikzpicture}\nonumber\\
	&\equiv\,\,\, -\int d^4x V^f_{\rm eff} \quad\,\,\,-\int d^4x V^s_{\rm eff}.
\end{align}

The expression for the first term can be obtained from (\ref{SunSetFull}) with the simplifications detailed above to find
\begin{widetext}
 \begin{align}
    -V^s_{\rm eff}&	=\frac{1}{12}(\mu^2)^{2\epsilon}\int\frac{d^dq d^dk}{(2\pi)^{2d}}(\mathcal D^3 V)^{(abc)}{\mathcal C}^{\,d}_{a}(q){\mathcal C}(p)^{\,e}_b{\mathcal C}(p+q)^{\,f}_{c}(\mathcal D^3 V)_{(def)}\nonumber\\
	&=\frac{1}{12}\left[\mathcal I_{V''}(V''')^2+\frac{n \mathcal I_{V'',F'V'/F}}{3}\left( \left(\frac{F'V'}{F}\right)^{\prime}+2F^{\prime}\left(\frac{V'}{F}\right)^{'}\right)^2\right];
\end{align}
where $()$ stands for symmetrisation. The second term follows from eq.~(\ref{FlyFull}),
\begin{align}
    -V^f_{\rm eff}&
	=\frac18 (\mu^2)^{2\epsilon}\int \frac{d^dqd^dk}{(2\pi)^{2d}}G^{aa'}G^{cc'}{\mathcal C}(q)^{\,\,b}_{a'}{\mathcal C}(k)_{c'}^{\,\,d}\left(\frac13 \mathcal R_{acbd} ((q-k))^2+{ ~((a\leftrightarrow c)+(a\leftrightarrow d))~}- (\mathcal D^4V)_{(abcd)} \right)\nonumber\\
	&=-\frac{1}{6}\left[n\mathcal R_h\mathcal  J_{V''}\mathcal J_{F'V'/F}\left(V''+\frac{F'V'}{F}\right)+n(n-1)\mathcal R_\varphi \mathcal J_{F'V'/F}^2\left(\frac{F'V'}{F}\right)\right]\nonumber\\
	&-\frac18\Bigg[\mathcal J_{V''}^2(V'''')+\frac{n \mathcal J_{V''}\mathcal J_{F'V'/F}}{3}\left\{\left(\frac{F'V'}{F}\right)''+\frac{F'}{F}[3V'''-4F'\left(\frac{V'}{F}\right)']-2\frac{F'}{F}\left(\frac{F'V'}{F}\right)'+2\left(F'\left(\frac{V'}{F}\right)'\right)'\right\} \nonumber\\
	&\qquad\qquad+\frac{n^2+2n}{3}\mathcal J_{F'V'/F}^2\left(\frac{F'}{F}\left(\frac{F'V'}{F}\right)'+2\frac{F^{\prime,2}}{F}\left(\frac{V'}{F}\right)'\right)\Bigg]\,.
\end{align}
\end{widetext}
where in the index exchange for the curvature term the momentum should be changed accordingly.

The functions $\mathcal I_{\alpha,\beta}$ and $\mathcal J_\alpha$ are the solutions to loop integrals, as outlined in sec.~\ref{secTH} it is convenient to use dimensional regularization in $d=4-2\epsilon$, in which case,
\begin{align}\nonumber\label{loopint}
\mathcal I_{\alpha,\beta}&=(\mu^2)^{2\epsilon}\int\frac{d^dq d^dk}{(2\pi)^{2d}}\frac{1}{(q^2+\alpha)((k+q)^2+\alpha)(k^2+\beta)}\\\
\mathcal J_{\alpha}&=(\mu^2)^\epsilon\int \frac{d^dq}{(2\pi)^d}\frac{1}{q^2+\alpha}
\end{align} 
with $\mu$ the renormalisation scale and an explicit form is drawn from \cite{Ford:1992pn} and shown in Appendix~\ref{app}, see ref.~\cite{Ford:1991hw} for a partial differential equation derivation of these integrals.

Before summarising we comment on checks to our computation. A self-consistency check is the cancellation of non-1PI terms which we have checked explicitly at the 2 and 3 loop order. A further check is the derivation of the symmetry factors from our Gaussian integral formulae, and the agreement with \cite{Palmer:2001vq}. Finally, in the flat case with $N=4$ $(n=3)$, eqs.~(\ref{SunSetFull},\ref{FlyFull}) yield the 2 loop self-corrections to the SM potential and they do indeed reproduce the results of ref~\cite{Ford:1992pn} eq.~(5.2) prior to renormalisation \footnote{Which in practice means substituting the functions $\hat I, \hat J$ for $I, J$}.


\section{Summary}\label{Summ}

This letter presented a covariant procedure for arbitrary loop correction computation on a general scalar manifold with gauge interactions and gravity via functional methods and an expansion on field derivatives, defining in the process an invariant partition function and covariant correlation functions off-shell which we connected to the geometric LSZ reduction formula. The formula for the effective action is given in eq.~(\ref{CR}) together with the steps for the computational procedure and applies to an arbitrary tree level action. In the derivation, we have neglected terms proportional to the source and $n\geq 2$ powers of the field which do not affect the $S$-matrix and arise when expanding around field-points away from the vacuum; we nevertheless believe their merit further study, deferred to future work. The procedure was put to use in the computation of two loop contributions to the effective potential of an $O(N)$ symmetric $N$-dimensional manifold with an arbitrary (but $O(N)$ symmetric) potential, readily applicable in HEFT with $N=4$, and agreeing in the appropriate limit with results from~\cite{Ford:1992pn}.


\section{Acknowledgements}

The authors would like to thank Elizabeth E. Jenkins, and Aneesh V. Manohar for helpful comments on the draft. R. A. and M. W. are supported by the STFC under Grant No. ST/T001011/1.


\bibliography{LibrarY.bib}

\appendix

\section{}\label{app}
Following \cite{Ford:1992pn}, we present in this appendix the solutions to integrals $\mathcal I_{\alpha,\beta}, \mathcal J_\alpha$ defined in eq.~(\ref{loopint}). We use dimensional regularisation with $d=4-2\epsilon$, and define the renormalisation scale to be $\mu$. Here, we have opted to leave formulae in terms of bare parameters and hence have left the UV divergences explicit as poles in $\epsilon$, which will be cancelled by renormalising in your scheme of choice. The finite remainder is the relevant piece for the effective potential.
We begin with the solution to $\mathcal J_\alpha$ as:
\begin{align}
    \mathcal J_\alpha&=\frac{(\mu^2)^\epsilon}{(2\pi)^d}\int\frac{d^dk}{k^2+\alpha}\nonumber\\
    &=\frac{(\mu^2)^\epsilon}{(4\pi)^{d/2}}\Gamma(1-d/2)\alpha^{d/2-1}
\end{align}
where $\Gamma$ is the usual Gamma function. Recall in the two-loop potential computation from sec.~\ref{TwoLVeff}, $\mathcal J_\alpha$ always appears with a $\mathcal J_\beta$. Hence, the Laurent series in $\epsilon$ we need to construct is:

\begin{align}
    \mathcal J_\alpha \mathcal J_\beta &=\frac{(\mu^2)^{2\epsilon}}{(4\pi)^{d}} \Gamma(1-d/2)^2 (\alpha\beta)^{d/2-1}\nonumber\\
    &=\mathcal T^2 \frac{\alpha\beta}{(4\pi)^4}\bigg[\frac{1}{\epsilon^2}+\frac{1}{\epsilon}\left\{2-\log\Big(\frac{\alpha}{\mu^2}\Big)-\log\Big(\frac{\beta}{\mu^2}\Big)\right\}\nonumber\\
    &+\bigg\{3+\frac{\pi^2}{6}-2\log\Big(\frac{\alpha}{\mu^2}\Big)-2\log\Big(\frac{\beta}{\mu^2}\Big)\nonumber\\
    &+\frac{1}{2}\left(\log\Big(\frac{\alpha}{\mu^2}\Big)+\log\Big(\frac{\beta}{\mu^2}\Big)\right)^2\bigg\}+\mathcal O(\epsilon)\bigg],
\end{align}
with $\mathcal T=e^{-\epsilon \gamma}/(4\pi)^\epsilon$ and $\gamma$ Euler's constant. Note in $\overline{\rm{MS}}$, $\mathcal T=1$. Then $\mathcal I_{\alpha,\beta}$:

\begin{align}
    \mathcal I_{\alpha,\beta}&=(\mu^2)^{2\epsilon}\int\frac{d^dq d^dk}{(2\pi)^{2d}}\frac{1}{(q^2+\alpha)((k+q)^2+\alpha)(k^2+\beta)}\nonumber\\
    &=-\mathcal T^2\frac{\Delta +2}{2(4\pi)^4}\alpha\Big(\frac{1}{\epsilon^2}+\frac{a_1}{\epsilon}+a_2+\mathcal O(\epsilon)\Big),
\end{align}
where $\Delta=\beta/\alpha$, and
\begin{align}
        a_1&=3-\frac{2\Delta \log(\Delta)}{\Delta+2}-2\log\Big(\frac{\alpha}{\mu^2}\Big)\\
        a_2&=7+\frac{\Delta( \log^2(\Delta) -6\log(\Delta))}{\Delta+2}-2\log^2\Big(\frac{\alpha}{\mu^2}\Big)\nonumber\\
        &\qquad-2a_1\log\Big(\frac{\alpha}{\mu^2}\Big)+\zeta(2)+8\xi(\Delta).
\end{align}
The function $\xi$ is defined, if $\Delta<4$, as:
\begin{align}
    \xi(\Delta)&=\frac{\sqrt{\Delta(4-\Delta)}}{\Delta+2}\int^\theta_0 \log(2\sin a)da,
\end{align}
with $\sin \theta = \sqrt{\Delta}/2$, and for $\Delta>4$ as:
\begin{align}
    \xi(\Delta)&=\frac{\sqrt{\Delta(\Delta-4)}}{\Delta+2}\int^\alpha_0 \log(2\cosh a)da,
\end{align}
where $\cosh \alpha =\sqrt \Delta/2$. Further, the function $\xi(\Delta)$ is continuous for $\Delta=4$ and either expression can be used.

\end{document}